\documentclass[twocolumn,aps]{revtex4}
\usepackage{graphicx}
\usepackage{bm}
\begin {document}

\title{A Dynamical Study of the Disordered Quantum $p=2$ Spherical Model}

\author{Michal Rokni}
\author{Premala Chandra}
\altaffiliation[Present address: ]{Department of Physics, Rutgers
University, Piscataway, NJ 08855, USA}

\affiliation{Chandra Tech Consulting LLC, 123 Harper Street, Highland
Park, NJ 08904}

\begin{abstract}
We present a dynamical study of the disordered quantum $p=2$ spherical
model at long times. Its phase behavior as a function of spin-bath
coupling, strength of quantum fluctuations and temperature is
characterized, and we identify different paramagnetic and coarsened
regions. A quantum critical point at zero temperature in the limit of
vanishing dissipation is also found. Furthermore we show analytically
that the fluctuation-dissipation theorem is obeyed in the stationary
regime.
\end{abstract}

\maketitle

\section{Introduction}
\label{intro}

The dynamics of quantum systems far from equilibrium present many
conceptual challenges, particularly in the area of quantum glasses. The
classical analogues of these systems display broad relaxation spectra;
furthermore their long-time behavior depends explicitly on details of
sample history. In this paper a motivating question is whether the
introduction of quantum fluctuations in a classical system out of
equilibrium results in qualitatively new behavior at long time-scales.
To date, the introduction of quantum fluctuations in theoretical glassy
models has not led to qualitative changes, e.g., new slow quantum
modes, in their long-time dynamics
\cite{CuL98,CuL99,BiP02,KeC01,KCY01}. Quantum fluctuations often drive
the glass transition to be first-order
\cite{CGS00,CGS01,BiC01,WSW02,NiR98}, and thus only qualitatively
affect the dynamical behavior on {\em short}-time scales. However there
are examples where, in the absence of dissipation, quantum fluctuations
lead to a quantum critical point at zero temperature; examples include
the periodic Josephson array \cite{KFI99} and infinite-component
quantum rotors \cite{YSR93, Kop94}. Detailed characterization of the
phase space as a function of dissipation, temperature, and quantum
fluctuations remains to be studied in both cases.

Careful experimental studies of the dipolar magnet ${\rm LiHo}_{x}{\rm
Y}_{1-x}{\rm F}_{4}$ indicate that the current theoretical picture of
quantum glassy systems is not the full story \cite{AeR98}. An
appropriate model for the description of the magnetic properties of
this material consists of Ising spins, representing the states of the
magnetic ions, residing on random sites, and interacting via
short-range ferromagnetic couplings and long-range dipole-dipole
interactions. The sign of the latter interaction depends on the
relative position of the spins, leading to the possibility of competing
interactions and to frustration. For $x = 1$, the short-range component
of the interaction dominates and the system is a ferromagnet. However,
the Curie temperature decreases as $x$ is reduced. The effects of
quantum mechanics can be tuned in this system by application of a
magnetic field transverse to the spins. Even in the disordered
ferromagnetic state ($x \sim 0.4 $), there are qualitative effects of
the transverse field, particularly near the observed quantum critical
point, that have not been described by theoretical models
\cite{AeR98,BBR99,BRA01}.

Motivated by these experiments, we study the dynamics of a long-range
quantum disordered ferromagnet, the quantum spherical model with
Gaussian random two-spin interactions ($p=2$). Classically the statics
of this model are trivial, as the free energy has only two minima, and
it can be solved exactly with a replica symmetric ansatz \cite{KTJ76}.
However, for almost any initial conditions, the dynamics of this
classical system are out-of equilibrium with power-law decays in many
of the observed quantities \cite{CuD95a}. The origin of this slow
relaxation is very different from that of classical mean-field spin
glasses. In these latter systems the time evolution is determined by
the complex structure of the phase space \cite{Cug02}. By contrast the
free energy landscape of the $p=2$ case is very simple. However,
statistically the overlap between the initial and the equilibrium
configurations is very small. This is qualitatively similar to the
situation in a weakly disordered ferromagnet after a rapid temperature
quench where the resulting domain structure bears little resemblance to
the uniform equilibrium state. For this reason we refer to the
non-ergodic phases of the $p=2$ spherical model as coarsened.

Naturally the infinite-range nature of this Hamiltonian implies the
absence of any length-scale in real-space. The spherical character of
the spins combined with the quadratic interaction lead to reduced
nonlinearities in this disordered $p=2$ model, even compared to other
mean-field spin glasses. These two key features, specifically the
infinite-range coupling and the minimal nonlinearities, make this model
amenable to analytic treatment, and it can serve as a testing ground
for necessary assumptions made in more complicated glassy systems. The
disordered quantum $p=2$ spherical model bears a strong resemblance to
that of the infinite-range $M$-component quantum rotors in the limit $M
\rightarrow \infty$ \cite{YSR93,Kop94}. More explicitly, the
$M$-component quantum rotor Hamiltonian is composed of $M$ replicas
which each individually resemble the $p=2$ system, where the
inter-replica coupling scales inversely with $M$ \cite{Kop94}. Here we
extend previous studies of these and related rotor Hamiltonians
\cite{YSR93,Kop94,KCY01} by performing a detailed characterization of
the phase behavior of the disordered quantum $p=2$ spherical model,
noting that our results recover those previously obtained in particular
well-defined limits.

In this study we characterize the phase behavior of the disordered
$p=2$ spherical model as a function of quantum and thermal fluctuations
and the spin-bath coupling; we identify its distinct paramagnetic and
coarsened regions by studying its long-time dynamical behavior. Known
results are reproduced in both the high- and the zero- temperature
limits. We also demonstrate analytically that the
fluctuation-dissipation theorem remains valid in the stationary regime.

The plan of the paper is as follows. In Section \ref{FotP} we present
the model. Its dynamical equations, which we derive using the Keldysh
formalism, are discussed in Section \ref{DEqs}. The stationary parts of
the retarded and the statistical Green functions, that we find in
Section \ref{Stationary}, are then shown analytically to obey the
fluctuation-dissipation theorem. In Section \ref{EA} an expression for
the Edwards-Anderson order parameter is presented with the resulting
emergence of a quantum critical point. Details of the phase diagram are
discussed in Section \ref{PD} for all the coarsened and the
paramagnetic regions. We end with a discussion, Section
\ref{Discussion}, where, after summarizing the main features of this
work, we mention several open questions that emerge from our study.

\section{Formulation of the Problem}
\label{FotP}

In this paper we study the dynamics of the disordered quantum $p=2$
spherical spin glass model coupled to an Ohmic bath. There are $N$
spins that can be written in vector form as ${\bf
s}=(s_{1},s_{2},\ldots,s_{N})$, and we will assume that $N\rightarrow
\infty$. The Hamiltonian of the system takes the form
\begin{equation}
  \label{h}
  {\cal H} = {\cal H}_{0} + {\cal H}_{\rm SS} + {\cal H}_{\rm B}
           + {\cal H}_{\rm SB},
\end{equation}
where ${\cal H}_{0}$, ${\cal H}_{\rm SS}$, ${\cal H}_{\rm B}$ and
${\cal H}_{\rm SB}$ are the free, the spin-spin, the bath, and the
spin-bath contributions respectively.

In the absence of interactions the free Hamiltonian of the spin system
is given by
\begin{equation}
  \label{h0}
  {\cal H}_{0} = \frac{{\bm \pi}^{2}}{2m},
\end{equation}
where the conjugated momenta $\bm{\pi}=(\pi_{1},\pi_{2},\ldots,
\pi_{N})$ obey the usual commutation relations
$[s_{i},s_{j}]=[\pi_{i},\pi_{j}]=0$, and $[\pi_{i},s_{j}]=-i
\delta_{ij}$; here we use a unit system where $\hbar=k_{B}=1$. The
``mass'' $m$ is a constant that introduces dynamics into the problem.
As we will explain below, the dynamics of the system can be tuned from
the classical to the quantum regimes by varying the mass $m$. We note
that there are other methods of quantization (e.g. application of a
transverse field) besides this "kinetic" one; if done correctly, they
should all yield the same results \cite{NiR98}.

The spins interact with each other via a spin-spin interaction which,
for the $p=2$ model, is given by
\begin{equation}
  \label{spin-int}
  {\cal H}_{\rm SS}[J] = \sum_{i_{1}<i_{2}}^{N} J_{i_{1}i_{2}} s_{i_{1}}
  s_{i_{2}},
\end{equation}
with randomly distributed matrix elements $J_{ij}$, that obey
$\overline{J_{ij}}=0$, $\left. \overline{J_{ij} J_{kl}} \right|_{i \neq
k \, {\rm or} \, j \neq l} = 0$, and $\overline{J_{ij}^{2}} = J^{2}/N
$, where $\overline{\cdots}$ represents averaging over the random
spin-spin interaction.

In the spherical model the spins, $s_{i}$, are constrained to a
spherical surface such that $ \sum_{i=1}^{N} \overline{ \left \langle
s_{i}^{2} \right \rangle} = N$, where $\langle \cdots \rangle$
represent quantum averaging. Thus $s_{i}$ are continuous variables,
that behave like coordinates. However, since this model has been
introduced as an analytically accessible spin problem, we will keep the
notation and the nomenclature. The spherical constraint is initially
introduced into the Lagrangian, which results in the additional term,
$\frac{1}{2} \, z(t) \, {\bf s}^{2}$, in the Hamiltonian. The
time-dependent function, $z(t)$, is a Lagrange multiplier. Thus the
free Hamiltonian ${\cal H}_{0}$ (\ref{h0}), is effectively replaced by
\begin{equation}
  \label{new-h0}
  {\cal H}_{0}=\frac{{\bm \pi}^{2}}{2m}+\frac{1}{2} \, z(t) \, {\bf
  s}^{2}.
\end{equation}

Each spin in the system is connected to a system of $N_{\rm B}$
independent harmonic oscillators that constitute the bath. The free
Hamiltonian of the bath is given by
\begin{equation}
  \label{hB}
  {\cal H}_{\rm B} = \frac{1}{2} \sum_{a=1}^{N_{\rm B}}
                 \left[ \frac{{\bf P}_{a}^{2}}{M_{a}}
                 + M_{a} \omega_{a}^{2} {\bf x}_{a}^{2}
  \right],
\end{equation}
where ${\bf x}_{a}=(x_{a1},x_{a2},\ldots ,x_{aN})$ are the oscillator
coordinates of mode $a$, and $x_{ai}$ is the coordinate of the
oscillator that interacts with spin $s_{i}$. The conjugate momenta
${\bf P}_{a}=(P_{a1},P_{a2},...P_{aN})$ are related to ${\bf x}_{a}$
through the commutation relations $[x_{ai},x_{aj}]=[P_{ai},P_{aj}]=0$,
and $[P_{ai},x_{aj}]=-i \delta_{ij}$. The mass and frequency of mode
$a$ of the oscillators are given by $M_{a}$ and $\omega_{a}$,
respectively. For simplicity we assume a linear interaction between the
spins and the bath \cite{LCD87}
\begin{equation}
  \label{spin-b-int}
  {\cal H}_{\rm SB}[C_{a}]=\sum_{a=1}^{N_{\rm B}}\, C_{a} {\bf x}_{a}
  \cdot {\bf s},
\end{equation}
where the coupling coefficients are given by $C_{a}$. We are aware that
this model for a bath can lead to unwanted divergencies due to
low-frequency contributions in the absence of an infrared cut-off
\cite{Ler-pc}. However, this problem does not arise in our case due to
the structure of the relevant spectral density, as we discuss later in
the paper. Such a simple interaction between the spin system and the
bath allows us to solve the problem analytically.

The Hamiltonian for the spin system is actually that of a harmonic
oscillator. Because the coupling to the bath (\ref{spin-b-int}) is
linear in the spin operator, it does not alter this statement. In the
paramagnetic phase the spin-spin interaction is negligible, and we
recognize that $\sqrt{z(t)/m}$ is the eigenenergy of the oscillator. We
are interested in the long-time behavior of $z(t)$ when it saturates to
a constant $z_{\infty}$. In order to understand whether the dynamics of
the spin system are governed by classical or quantum fluctuations, we
compare the important energy-scale of the system to the temperature
$T$. At constant $T$ when $m$ is increased, the eigenenergy of the spin
Hamiltonian decreases. Once the dominant energy-scale is smaller than
$T$, the system dynamics are governed by classical behavior. Similarly,
when $m$ is decreased, $T$ becomes smaller than the dominant
energy-scale, and quantum dynamics become more important. As a
cautionary aside, we note that $z_{\infty}$ may depend on $m$ in the
paramagnetic regime; however we will show that to leading order it is
either independent of $m$ or proportional to $m^{-1}$, so that the
preceding heuristic argument is still valid.

In the coarsened phase the spin-spin coupling cannot be neglected.
After averaging over this interaction, it is harder to recognize the
eigenenergies. However, the results presented here indicate that when
the coupling to the bath is infinitesimal, the signatory parameter is
$x=\sqrt{mT^2/J}$: for $x$ large, the dynamics are dominated by
classical effects whereas for $x$ small quantum effects are involved.
It is only at $T=0$ that the dynamics are purely quantum in nature. We
conclude therefore that the energy-scales $T$ and $\sqrt{J/m}$ should
be compared in order to find whether the dynamics are purely classical
or partially quantum (except at $T=0$). This is consistent with the
previous discussion of the paramagnetic phase, where small mass
corresponds to a regime where quantum fluctuations could become
dominant.

\section{The Dyson Equations for the Spin Green Functions}
\label{DEqs} The dynamics of the spin system can be characterized by
Dyson equations associated with its Green functions. Here we use a
nonequilibrium diagrammatic description, the Keldysh technique
\cite{Kel65,LiP}, and take the initial conditions to be random. Two
Green functions, or rather three {\em real} functions, are needed to
describe the system, they are the retarded and the statistical (or
Keldysh) Green functions, that are defined by
\begin{equation}
  \label{Grt}
  G^{\rm r}(t,t')=-i \Theta(t-t')
                   \overline{\left \langle
                   s_{i}(t)s_{i}(t')-s_{i}(t')s_{i}(t)
                   \right \rangle},
\end{equation}
and
\begin{equation}
  \label{Gst}
  G^{\rm s}(t,t')=-i
                   \overline{\left \langle s_{i}(t)s_{i}(t')+s_{i}(t')s_{i}(t)
                   \right \rangle },
\end{equation}
respectively. For the $p=2$ spherical model the Green functions are
diagonal in the spin index, due to both the disorder averaging, and the
infinite-range nature of the interaction. The retarded Green function
is proportional to the ``response'' in the spin glass nomenclature,
while the statistical Green function is proportional to the
``correlation''. At equal times we demand that $G^{\rm s}(t,t)=-2 i $
due to the spherical constraint, thus setting $z(t)$.

When a system is at equilibrium the Green functions that describe it
are a function of the time difference only. In that case one can
Fourier transform $G^{\rm r}(t-t')$ and $G^{\rm s}(t-t')$ over the time
difference to obtain $\tilde{G}^{\rm r}(\omega)$ and $\tilde{G}^{\rm
s}(\omega)$. Furthermore, at equilibrium the retarded and the
statistical Green functions are related via the fluctuation-dissipation
theorem (FDT)
\begin{equation}
  \label{FDT}
  \tilde{G}^{\rm s}(\omega) =
  2i \coth{\left(\frac{\omega}{2T}\right)} {\rm Im}
  \tilde{G}^{\rm r}(\omega),
\end{equation}
where $T$ is the equilibrium temperature of the system. When the spin
system is in the paramagnetic phase, it is in equilibrium, and
therefore it obeys FDT. Although in the coarsened phase the system is
{\em not} in equilibrium, we will show that one can identify unique
stationary contributions to the retarded and the statistical Green
functions, that are related via the FDT, where the temperature is that
of the bath.

\begin{figure}
  \begin{center}
    \includegraphics[width=8.25cm]{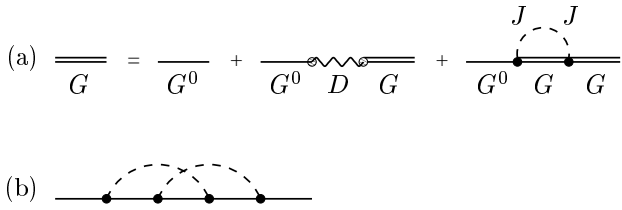}
  \end{center}
  \caption{\label{diagram}(a) Dyson equations depicted diagrammatically
  in a schematic manner, in the self-consistent-Born approximation. The dashed
  lines represent averaging over spin-spin interaction. (b)
  An example of a diagram with crossed spin-spin interaction averaging
  lines, that is proportional to $1/N$, and therefore is neglected in
  our calculation.}
\end{figure}

When the spin system is in the coarsened phase, the Green functions
depend on {\em both} times, and not only on the time difference. The
spin Green functions in the time domain obey the Dyson equations
\begin{eqnarray}
  \label{DE-Gr-t}
  \lefteqn{-\left[m \frac{\partial^{2}}{\partial t^{2}}+z(t) \right]
  G^{\rm r}(t,t')=} \\ \nonumber & & \qquad \qquad \qquad
  \delta \, (t-t') + \int {\rm d} t'' \,
  \Sigma^{\rm r}(t,t'')
  G^{\rm r}(t'',t'),
\end{eqnarray}
and
\begin{eqnarray}
  \label{DE-Gs-t}
  \lefteqn{-\left[m \frac{\partial^{2}}{\partial t^{2}}+z(t) \right]
  G^{\rm s}(t,t') =} \\ \nonumber & & \qquad
  \int {\rm d} t'' \left[
  \Sigma^{\rm s}(t,t'')
  G^{\rm a}(t'',t')
  + \Sigma^{\rm r}(t,t'')
  G^{\rm s}(t'',t') \right],
\end{eqnarray}
where $G^{\rm a}(t,t')=[G^{\rm r}(t',t)]^{*}$ is the advanced Green
function, and $\Sigma^{\rm r/s}(t,t')$ are the self-energy functions.
Note that the Dyson equations above apply to the system in the
paramagnetic phase also, where $z(t)$ is a constant and all other
functions depend only on the time difference. Assuming $N$ is large, we
consider the self-energy functions in the self-consistent-Born
approximation [see Fig.\ \ref{diagram}(a)], in which case they are
given by
\begin{equation}
  \label{self-energy}
  \Sigma^{\rm r/s}(t,t')=D^{\rm r/s}(t-t') + J^{2}G^{\rm r/s}(t,t').
\end{equation}
The bath contributions to the self-energy, which include the spin-bath
interaction vertices, are $D^{\rm r/s}(t-t')$. We have neglected
``crossing'' diagrams [e.g., see Fig.~\ref{diagram}(b)], since they are
at least a factor of $1/N$ smaller than the diagrams considered. This
approximation is exact in the $N \rightarrow \infty$ limit.

We remark that the equation for the retarded Green function
(\ref{DE-Gr-t}) is completely decoupled from the statistical Green
function. This is a result of two model-dependant features: (i) that
$p=2$, (ii) that the coupling to the bath is linear in the spin
operators. In a $p \geq 3$ model, or in any $p$ model with a nonlinear
coupling to the bath, the equation for the retarded Green function
would include the statistical Green function as well.

The detailed nature of the relaxation of a spin system far from
equilibrium depends on the time it has spent in its low-temperature
phase before the start of the measurement. This phenomenon is known as
aging. This dynamical behavior has been described using scaling and
other phenomenological approaches \cite{BCK,Cug02}. More recently aging
has been studied analytically in microscopic models where mean-field
models are exact \cite{BCK,Cug02}. It was realized that in these
infinite-range systems the time evolution is dominated by motion in
flat directions in phase space rather than by transitions over
barriers; with random initial conditions, the system continues to
evolve and never approaches its local equilibrium. For this
``weak-ergodicity'' breaking scenario \cite{BCK,Cug02,Bou92,CuK9395},
it is crucial that the dimensionality of phase space is very large.
Because the spin system is out of equilibrium, its memory of its
initial states will depend specifically on the time when the
measurement commenced, $t'$, and the time when it ended, $t=t'+ \Delta
t$, and not simply the time difference, $\Delta t=t-t'$. In the
weak-ergodicity breaking scenario, if the waiting-time $t'$ is finite,
the system can evolve in the long-time limit so as to lose complete
memory of its initial configuration at $t'$. This assumption of weak
long-term memory has been shown to be self-consistent for several
classical long-range glassy and coarsened models
\cite{BCK,Cug02,Bou92,CuK9395}. In the quantum case, it is expected to
hold in general \cite{CuL98,CuL99}, though it may be violated near a
quantum critical point where the decoherence time diverges.

We assume that the weak-ergodicity breaking scenario applies to our
system, and look for a solution that obeys it. More specifically,
according to the weak-ergodicity breaking scenario, one expects the
statistical Green function to decay in the following manner
\begin{eqnarray}
  \label{Gs-t'->inf}
  \lim_{t' \rightarrow \infty} G^{\rm s}(t,t')& = &
  -2 i q + G^{\rm s}_{\rm ST}(t-t'), \\
  \label{Gs-t-t'->inf}
  \lim_{t-t' \rightarrow \infty} G^{\rm s}_{\rm ST}(t-t') & = & 0,
  \\
  \label{Gs-t->inf}
  \lim_{t \rightarrow \infty}G^{\rm s} (t, t') & = & 0,
\end{eqnarray}
for $t \geq t'$. The Edwards-Anderson parameter,
$q=-\lim_{t-t'\rightarrow \infty} \lim_{t' \rightarrow \infty}G^{\rm
s}(t,t')/2i$, characterizes the coarsened phase. For large waiting
times $t'$, the statistical Green function is given by a term that is
proportional to $q$, plus a stationary function $G^{\rm s}_{\rm
ST}(t-t')$ that decays to zero at infinity. This serves as a definition
of $G^{\rm s}_{\rm ST}(t-t')$.

Using the definition for $G^{\rm s}_{\rm ST}(t-t')$, the statistical
Green function for any times $t'$ and $t$ can be written as a sum of
$G^{\rm s}_{\rm ST}(t-t')$ and an aging function $G^{\rm s}_{\rm
AG}(t,t')$, that is not time translationally invariant,
\begin{equation}
  \label{aging}
  G^{\rm s}(t,t')= G^{\rm s}_{\rm AG}(t,t')+G^{\rm s}_{\rm ST}(t-t').
\end{equation}
At equal times $G^{\rm s}(t,t)=-2i$ due to the spherical constraint, so
that Eq.~(\ref{Gs-t'->inf}) at equal times yields
\begin{equation}
  \label{equal-t}
  \lim_{t \rightarrow \infty} G^{\rm s}_{\rm AG}(t,t)=-2iq \;\; \mbox{and} \;\;
  G^{\rm s}_{\rm ST}(0)=-2i(1-q).
\end{equation}

In order to write the Dyson equations (\ref{DE-Gr-t})-(\ref{DE-Gs-t}),
we need the contribution of the bath to the self-energy [see
(\ref{self-energy})]. In the frequency domain the free-bath Green
functions for mode $a$, that are associated with the free-bath
Hamiltonian ${\cal H}_{B}$, are given by
\begin{equation}
  \label{Da-r-w}
  \tilde{D}^{\rm r}_{a}(\omega) = \frac{1}{\omega-\omega_{a}+i \delta/2}-
                        \frac{1}{\omega+\omega_{a}+i \delta/2},
\end{equation}
\begin{equation}
  \label{Da-s-w}
  \tilde{D}^{\rm s}_{a}(\omega) = -2 \pi i \coth{\left(\frac{\omega}{2T}\right)}
  \left[\delta(\omega-\omega_{a}) + \delta(\omega+\omega_{a}) \right],
\end{equation}
where $T$ is the bath temperature. Fourier transforming into the time
domain one obtains
\begin{eqnarray}
  \label{Da-t}
  D^{\rm r}_{a}(t) & = & -2 \Theta (t) \sin{\omega_{a}t}, \\
  D^{\rm s}_{a}(t) & = & -2 i \coth{\left(\frac{\omega_{a}}{2T}\right)}
  \cos{\omega_{a} t}.
\end{eqnarray}

We assume that the bath is in equilibrium, namely that it is so large
that it is unaffected by its interaction with the spin system.
Therefore we take the bath Green functions to be those in
(\ref{Da-r-w}) and (\ref{Da-s-w}), that depend only on the time
difference. Thus the contributions of the bath to the self-energy are
also time-translationally invariant. Furthermore, they are linear in
the bath Green functions. They are obtained by multiplying the Green
function of each mode by its density of states, $(2 M_{a}
\omega_{a})^{-1}$, and by the spin-bath coupling constant squared
$C_{a}^{2}$, and summing over all modes
\begin{equation}
  \label{Drt}
  D^{\rm r}(t)=- 2 \Theta(t) \int_{-\infty}^{\infty} {\rm d} \omega \,
  I(\omega)
  \sin {\omega t},
\end{equation}
for the retarded self-energy, and
\begin{equation}
  \label{Dst}
  D^{\rm s}(t)=- 2 i \int_{-\infty}^{\infty} {\rm d} \omega \,
  I(\omega) \coth{\left(\frac{\omega}{2T}\right)}
  \cos {(\omega t)},
\end{equation}
for the statistical self-energy, where \cite{LCD87}
\begin{equation}
  I(\omega)=\sum_{a=1}^{N_{b}}\frac{C_{a}^{2}}{2 M_{a} \omega_{a}}
  \; \delta (\omega-\omega_{a}).
\end{equation}
The bath temperature $T$ enters the equations for the spin Green
functions only through the bath contribution to the statistical
self-energy.

In the case of an Ohmic bath $I(\omega)=(\gamma \omega/\pi)
\exp{(-|\omega|/\Omega)}$ \cite{LCD87}, where $\gamma$ is a
dimensionless coupling constant, and $\Omega$ is an ultraviolet cutoff
that will be taken to infinity later on. Thus for the case of an Ohmic
bath
\begin{equation}
  \label{Drt-O}
  D^{\rm r}(t)=\frac{4 \gamma}{\pi} \Theta(t) \frac{\rm d}{{\rm d}t}
  \frac{\Omega}{1+ (\Omega t)^{2}}.
\end{equation}
When $\Omega \rightarrow \infty$ one can Fourier transform
(\ref{Drt-O}) and obtain
\begin{equation}
  \label{Drw}
  \tilde{D}^{\rm r}(\omega)= -2 \gamma \left(\frac{2 \Omega}{\pi}+ i \omega
  \right) \equiv \tilde{D}^{\rm r}(0) + \tilde{d}^{\rm r}(\omega).
\end{equation}
It would seem that since $\tilde{D}^{\rm r}(0) \propto \Omega$, the
limit $\Omega \rightarrow \infty$ is problematic. However, as we will
show, $\tilde{D}^{\rm r}(0)$ disappears from the equations for the spin
Green functions. Since the bath is in equilibrium it must obey the
fluctuation-dissipation theorem, so that $\tilde{D}^{\rm s}(\omega)= 2
i \coth{(\omega/2 T)}{\rm Im}\tilde{D}^{\rm r}(\omega)$. Thus
\begin{equation}
  \label{Dsw}
  \tilde{D}^{\rm s}(\omega)= -4 i \gamma \omega
  \coth{\left(\frac{\omega}{2T}\right)}.
\end{equation}
This result can also be obtained from (\ref{Dst}) with the Ohmic
density of states, directly.

\section{Stationary Regime}
\label{Stationary}

In this section we treat the retarded Green function $G^{\rm r}(t,t')$
at large times $t$ and $t'$, such that $t,t' \rightarrow \infty$, while
$t-t'$ is finite. We show that it is stationary, and that it is related
to the stationary part of the statistical Green function via the
fluctuation-dissipation theorem.

The Lagrange multiplier $z(t)$ is simply a function of one time
variable, therefore for large $t$ it should saturate to a constant
value $z_{\infty}=\lim_{t \rightarrow \infty} z(t)$. We assume that in
the coarsened phase $z_{\infty}$ is independent of temperature, which
indeed turns out to be consistent with our findings. Since $D^{\rm r}$
is independent of temperature, from Eq.~(\ref{DE-Gr-t}) it is evident
that once $z(t)$ saturates to $z_{\infty}$, $G^{\rm r}(t,t')$ no longer
depends on temperature. This agrees with the classical result found in
\cite{CuD95a}. Since $G^{\rm r}(t,t')$ does not depend on $G^{\rm s}$,
which is expected to age, one can find a time-translational solution
for $G^{\rm r}$, once $z_{\infty}$ is approached. This is a peculiarity
of the $p=2$ model, with the linear spin-bath coupling, since the
retarded Green function is independent of the statistical Green
function. We note that, consistent with our present discussion, no
aging was found in the response of the infinite-range $M$-component
rotors in the limit $M \rightarrow \infty$ \cite{KCY01}.

The time-translational invariance of $G^{\rm r}$ for large times can be
understood from a heuristic point of view. Because the response
function is decoupled from its statistical counterpart, it is
independent of $T$, and therefore must behave in the same manner for
all temperatures. At very large $T$, the system is paramagnetic and
thus the response is time-translationally invariant. Thus it must
exhibit this behavior at all temperatures. By contrast, the statistical
Green function, that is proportional to the spin correlation, is
temperature-dependent so that the above argument does not apply to it.

Assuming large enough times so that $z(t)$ has saturated, and therefore
$G^{\rm r}(t,t')=G^{\rm r}(t-t')$, one can Fourier transform
Eq.~(\ref{DE-Gr-t}) over $t-t'$ to obtain
\begin{equation}
  \label{DE-Gr-w}
  \tilde{G}^{\rm r}(\omega) =
  \frac{1}
       {m \omega^{2} - \bar{z}- \tilde{d}^{\rm r}(\omega) -
       J^{2}\tilde{G}^{\rm r}(\omega)},
\end{equation}
where $\bar{z}=z_{\infty}+\tilde{D}^{\rm r}(0)$. This is a quadratic
equation for $\tilde{G}^{\rm r}(\omega)$ which also applies to the
paramagnetic regime. However, $\bar{z}$ is still unknown, and will be
obtained from the equation for the statistical Green function. It will
be shown that $\bar{z}$ is finite, so that the divergence of
$\tilde{D}^{\rm r}(0)$ for $\Omega \rightarrow \infty$ does not pose a
problem.

In order to obtain an equation for $G^{\rm s}_{\rm ST}$, we write the
statistical self-energy in terms of the stationary and aging
contributions, $\Sigma^{\rm s}(t,t')=\Sigma^{\rm s}_{\rm
ST}(t-t')+\Sigma^{\rm s}_{\rm AG}(t,t')$, where
\begin{eqnarray}
  \label{se-ST}
  \Sigma^{\rm s}_{\rm ST}(t-t') & = &
  D^{\rm s}(t-t') + J^{2} G^{\rm s}_{\rm ST}(t-t'), \\
  \label{se-AG}
  \Sigma^{\rm s}_{\rm AG}(t,t') & = & J^{2} G^{\rm s}_{\rm AG}(t,t').
\end{eqnarray}
Assuming $t'$ is large, we substitute (\ref{Gs-t'->inf}) into
Eq.~(\ref{DE-Gs-t}) and obtain
\begin{widetext}
\begin{eqnarray}
  \label{DE-Gs-t1}
  \lefteqn{-\left(m \frac{\partial^2}{\partial t^{2}} + z_{\infty} \right)
  \left[- 2 i q + G^{\rm s}_{\rm ST}(t-t')\right]=} \nonumber \\
  & & \qquad \qquad \qquad \int_{-\infty}^{\infty} {\rm d}t''
      \left[ \Sigma^{\rm s}_{\rm ST}(t-t'')G^{\rm a}(t''-t') +
              \Sigma^{\rm r}(t-t'')G^{\rm s}_{\rm ST}(t''-t')\right] -
  2 i q \left[2 J^{2} \tilde{G}^{\rm r}(0) + \tilde{D}^{\rm r}(0)
  \right].
\end{eqnarray}
\end{widetext}
In the derivation of Eq.~(\ref{DE-Gs-t1}), one encounters terms of the
form $\int_{-\infty}^{\infty} {\rm d} t'' G^{\rm s}_{\rm AG}(t,t'')
F(t''-t')$, where $F$ can be any one of the stationary functions
$G^{{\rm r}/{\rm a}}$ or $D^{{\rm r}/{\rm a}}$. We note that in these
terms the aging function changes slowly compared to the stationary
ones. We exploit this feature and assume that $t''$ is close to $t'$,
since the retarded and advanced functions decay to zero when their
arguments become large. We therefore approximate
\begin{equation}
  \label{approx1}
  \int_{-\infty}^{\infty} {\rm d} t''
  G^{\rm s}_{\rm AG}(t,t'') F(t''-t') \simeq -2 i q \tilde{F}(0),
\end{equation}
where $\tilde{F}(\omega)$ is the Fourier transform of $F(t)$.
Similarly,
\begin{equation}
  \label{approx1'}
  \int_{-\infty}^{\infty} {\rm d} t''
  F(t-t'') G^{\rm s}_{\rm AG}(t'',t') \simeq -2 i q \tilde{F}(0).
\end{equation}
In addition, in obtaining Eq.~(\ref{DE-Gs-t1}), we used the fact that
the imaginary part of a retarded Green function at zero frequency is
zero.

Although we assumed that $t-t'$ is finite, we can Fourier transform
Eq.~(\ref{DE-Gs-t1}) since it applies to {\em any} $t-t'$ in the $t'
\rightarrow \infty$ limit. The Fourier transform of
Eq.~(\ref{DE-Gs-t1}) is
\begin{eqnarray}
  \label{DE-Gs-w}
  \lefteqn{\left\{m \omega^{2} - \bar{z} - \tilde{d}^{\rm r}(\omega) -
         J^{2} \left[\tilde{G}^{\rm r}(\omega)+\tilde{G}^{\rm a}(\omega)
               \right] \right\}
  \tilde{G}^{\rm s}_{ST}(\omega) =} \nonumber \\ & & \qquad
  \tilde{D}^{\rm s}(\omega) \tilde{G}^{\rm a}(\omega) -
  2 i q \, \delta(\omega)
  \left[2 J^{2} \tilde{G}^{\rm r}(0) + \bar{z}\right]. \qquad \qquad
\end{eqnarray}
Since $\lim_{t \rightarrow \infty} G^{\rm s}_{ST}(t) = 0$ the
expression for $\tilde{G}^{\rm s}_{ST}(\omega)$ cannot include terms
that are proportional to $\delta(\omega)$. We expect the other Green
functions in Eq.~(\ref{DE-Gs-w}) to be similarly well-behaved.
Therefore in the coarsened phase, since $q$ is finite, the term in the
square brackets that multiplies the delta function must be zero,
leading to the following equation for $\bar{z}$ and $\tilde{G}^{\rm
r}(0)$
\begin{equation}
  \label{z1}
  \bar{z} = - 2 J^{2} \tilde{G}^{\rm r}(0).
\end{equation}
In the paramagnetic phase $q=0$, and therefore Eq.~(\ref{z1}) is no
longer valid.

Another equation for $\bar{z}$ and $\tilde{G}^{\rm r}(0)$ is found by
substituting $\omega=0$ into Eq.~(\ref{DE-Gr-w}),
\begin{equation}
  \label{z2}
  \bar{z}= - \frac{1}{\tilde{G}^{\rm r}(0)} - J^{2} \tilde{G}^{\rm r}(0),
\end{equation}
which when compared with Eq.~(\ref{z1}) renders $[J \tilde{G}^{\rm
r}(0)]^{2} = 1$. Since $\tilde{G}^{\rm r}$ must be analytic in the
upper half plane, one obtains $\tilde{G}^{\rm r}(0)= -1/J$ and
therefore
\begin{equation}
\label{z-G}
  \bar{z} = 2 J,
\end{equation}
in the coarsened phase. Indeed as we anticipated, $\bar{z}$ is a
constant and does not depend on the temperature $T$. Furthermore,
$\bar{z}$ is finite, so the divergence of $\lim_{\Omega \rightarrow
\infty} \tilde{D}^{\rm r}(0)$ is not a problem; see the definition of
$\bar{z}$ following Eq.~(\ref{DE-Gr-w}), and the discussion following
(\ref{Drw}). The same $\bar{z}$ has been found for the classical $p=2$
model \cite{KTJ76}. In Ref.~\cite{CuL99} an equation for $z_{\infty}$
in terms of $\tilde{G}^{\rm r}$ and $\tilde{\Sigma}^{\rm r}$ was found
for the quantum disordered $p$-spin spherical models of $p \geq 3$;
there $\tilde{\Sigma}^{\rm r}$ depends on {\em both} $\tilde{G}^{\rm
s}$ and $\tilde{G}^{\rm r}$ so that a simple solution for $z_{\infty}$
does not emerge. However in the $p=2$ case with {\em linear} spin-bath
coupling $\tilde{\Sigma}^{\rm r}$ is {\em independent} of
$\tilde{G}^{\rm s}$. This decoupling leads to two distinct equations
for $z_{\infty}$ and $\tilde{G}^{\rm r}$, and thus to solutions for
both. More specifically (\ref{z1}), the analogue of the equation that
was found for $z_{\infty}$ in Ref.~\cite{CuL99}, contains {\em only}
$\tilde{G}^{\rm r}$; (\ref{DE-Gr-w}) provides a second equation for
$\bar{z}$ and $\tilde{G}^{\rm r}$, thereby permitting us to determine
$\bar{z}$.

Now that $\bar{z}$ is known, we can solve Eq.~(\ref{DE-Gr-w}) for
$\tilde{G}^{\rm r}(\omega)$ in the coarsened phase
\begin{widetext}
\begin{equation}
  \label{Gr-w}
  \tilde{G}^{\rm r}(\omega)=\frac{1}{2 J^{2}}
  \left\{
    m \omega^{2} - 2 J + 2 i \gamma \omega -
    \left[ \left(m \omega^{2}+ 2 i \gamma \omega \right)
           \left(m \omega^{2}-4 J + 2 i \gamma \omega \right) \right]^{1/2}
  \right\}.
\end{equation}
\end{widetext}
In solving the quadratic equation for $\tilde{G}^{\rm r}(\omega)$, we
choose the negative sign in front of the square root to ensure that
$\lim_{\omega \rightarrow \infty} \tilde{G}^{\rm r}(\omega)=0$.

We now demonstrate the FDT in the coarsened phase for the stationary
part of the Green functions. Using Eq.~(\ref{DE-Gr-w}) in
Eq.~(\ref{DE-Gs-w}), after the term with the delta function has been
eliminated, we obtain
\begin{equation}
  \tilde{G}^{\rm s}_{ST}(\omega) =
  \frac{\tilde{D}^{\rm s}(\omega) \left|\tilde{G}^{\rm r}(\omega) \right|^{2}}
       {1 - J^{2}\left|\tilde{G}^{\rm r}(\omega) \right|^{2}}.
\end{equation}
We write $\tilde{D}^{\rm s}(\omega)$ in terms of ${\rm Im}
{\tilde{D}^{\rm r}(\omega)}$ via the FDT, and taking the imaginary part
of Eq.~(\ref{DE-Gr-w}) we obtain
\begin{equation}
  \label{FDT-G}
  \tilde{G}^{\rm s}_{ST}(\omega) = 2 i \coth{\left(\frac{\omega}{2 T}\right)}
  {\rm Im}{\tilde{G}^{\rm r}(\omega)}.
\end{equation}
Thus, we have shown analytically that the stationary part of the
statistical Green function is related to the imaginary part of the
retarded Green function via the fluctuation-dissipation theorem, where
the temperature is that of the bath.

Now we have fully solved for the system's behavior in the stationary
regime. Once the retarded Green function has been found, (\ref{Gr-w}),
the stationary part of the statistical Green function is known as well
via the FDT, (\ref{FDT-G}).

\section{The Edwards-Anderson order parameter}
\label{EA} The Edwards-Anderson order parameter $q$ can be found by
substituting the expression that was obtained for the stationary part
of the statistical Green function, (\ref{FDT-G}) (FDT), into the
matching condition at zero time difference (\ref{equal-t}):
\begin{equation}
  \label{q}
  q=1 + \int_{-\infty}^{\infty} \frac{{\rm d} \omega}{2 \pi}
  \coth{\left(\frac{\omega}{2 T} \right)} {\rm Im}
  \tilde{G}^{\rm r} (\omega).
\end{equation}
The Edwards-Anderson parameter is in the range $0 \leq q \leq 1$. We
expect to obtain $q=0$ at equilibrium for the system in its
paramagnetic phase. We note that (\ref{q}) admits negative values of
$q$, which correspond to an instability of the coarsened phase; in this
case one should look for a paramagnetic solution.

As a check, expression (\ref{q}) for $q$ can be evaluated in two known
limits. The first is at high temperatures, when $T$ is much larger than
all energy-scales apart from $J$. In this case $\coth{\left( \omega/2 T
\right)} \simeq 2T/ \omega$ and we obtain
\begin{equation}
  \label{q-T>}
  q=1+ \frac{T}{\pi} \int_{-\infty}^{\infty}
  \frac{{\rm d}\omega}{\omega}
  {\rm Im} \tilde{G}^{\rm r}(\omega) = 1+T \, {\rm Re} \tilde{G}^{\rm r}(0)
  =1-\frac{T}{J},
\end{equation}
where we have used the Kramers-Kronig relation. This is the well-known
classical result \cite{KTJ76,CuD95a}, which should indeed be obtained
when $T$ is large. As an aside, we note that $T$ should be smaller than
$J$, since otherwise the system would undergo a phase transition into
the paramagnetic phase.

The other limit that we can check is $T \rightarrow 0 $, since $\lim_{T
\rightarrow 0} \coth{\left(\omega/2 T \right)} = {\rm sign}(\omega)$;
in this case
\begin{equation}
  \label{q-Eq-T=0}
  q = 1 +\int_{-\infty}^{\infty} \frac{{\rm d} \omega}{2 \pi}
  {\rm sign}{(\omega)} \, {\rm Im}\tilde{G}^{\rm r}(\omega).
\end{equation}
In equilibrium, $\int_{-\infty}^{\infty} {\rm d} \omega/(2 \pi) {\rm
sign}{(\omega)} {\rm Im} \tilde{G}^{\rm r}(\omega)=-1$ \cite{Mah} and
we obtain $q=0$, as expected.

In our calculation of $q$ we will use the following approximation
\begin{equation}
\label{approx}
  \coth{\left(\frac{\omega}{2T}\right)} \simeq
  \left\{
  \begin{array}{lll}
  {\rm sign}(\omega) & {\rm for} &
   2T < \omega, \\
  2 T /\omega & {\rm for} &
  2T > \omega, \\
  \end{array}
  \right.
\end{equation}
so that
\begin{equation}
  \label{q-approx}
  q \simeq 1 + \frac{2 T}{\pi} \int_{0}^{2T}
  \frac{{\rm d}\omega}{\omega} \,
  {\rm Im}\tilde{G}^{\rm r}(\omega) +
  \frac{1}{\pi} \int_{2 T}^{\infty}
  {\rm d}\omega \, {\rm Im}\tilde{G}^{\rm r}(\omega).
\end{equation}
We are aware that the model we are using for a bath can lead to
peculiar results since it does not incorporate an infrared cutoff for
the bath frequencies. In such a case the first integral in
Eq.~(\ref{q-approx}) might be finite even when $T \rightarrow 0$. We
have verified that this problem does not arise in our calculations,
since for $\omega \rightarrow 0$ ${\rm Im} \tilde{G}^{\rm r}(\omega)
\propto \omega^{\alpha}$, where $\alpha \geq 1$.

For a given average spin-spin interaction $J$ we look at the phase
space that is defined by the parameters $m,\gamma,$ and $T$. We focus
on the $\gamma \rightarrow 0$, and the $T=0$ planes. The coarsened
phase is defined by a finite $q$, while $q=0$ at any temperature
corresponds to the paramagnetic phase. We shall look for the $q=0$
line, knowing that beyond it (when $q<0$) the coarsened phase breaks
down, and the system must be in the paramagnetic phase. We have checked
numerically in the $\gamma \rightarrow 0$ and the $T=0$ planes, that
there is no coexistence of the paramagnetic and the coarsened phases,
except on the $q=0$ line. Furthermore, in section \ref{PM} we show that
the paramagnetic solution changes continuously into the coarsened one.
Thus this is a second order phase transition, and the $q=0$ line that
is found in the coarsened phase is indeed the phase boundary.

In the next section we describe the phase diagram. Here we concentrate
on finding the quantum critical point which resides in the $T=0$ plane
at the $\gamma \rightarrow 0$ limit. We expect the system to behave in
a coarsened manner for larger values of $m$. Upon decreasing $m$ we
anticipate that the quantum fluctuations will grow until the mass
reaches a critical value $m^{*}$, where the system undergoes a phase
transition into the paramagnetic state.

At this point we comment on the physical meaning of taking $\gamma
\rightarrow 0$. If one were to take $\gamma=0$ from the start, one
would initially decouple the spin system from the bath. In this case
the FDT relationship that we had found, (\ref{FDT-G}), that includes
the bath temperature, would not be valid---the spin system would not be
``aware'' of the bath's temperature unless it was coupled to it at some
point in time. By taking $\gamma \rightarrow 0$ we consider the spin
system to be coupled to the bath, with an infinitesimally small
coupling.

Substituting $\gamma \rightarrow 0$ into Eq.~(\ref{Gr-w}), we find
\begin{eqnarray}
  \label{Gr-w-B=0}
  \lefteqn{\left. {\rm Im} \tilde{G}^{\rm r}(\omega) \right|_{\gamma \rightarrow 0}
  = - \frac{1}{J} \left(\frac{m}{J} \right)^{1/2} \omega
  \left(1 - \frac{m \omega^{2}}{4 J} \right)^{1/2}} \nonumber \\
  & &  \qquad \qquad \qquad \qquad
  {\rm for} \;\;\; |\omega|<2 \left(\frac{J}{m} \right)^{1/2},
  \qquad
\end{eqnarray}
and $0$ otherwise. Substituting (\ref{Gr-w-B=0}) into
Eq.~(\ref{q-Eq-T=0}) for $T=0$, one can calculate $q$ exactly
\begin{equation}
  \label{q-B=T=0}
  q=1-\frac{4}{3 \pi \sqrt{m J}}.
\end{equation}
Taking $q=0$ in (\ref{q-B=T=0}) and solving for $m$ we find the quantum
critical point
\begin{equation}
  \label{qcp}
  m^{*}=\left(\frac{4}{3 \pi}\right)^{2} \frac{1}{J} \sim 0.18
  \frac{1}{J}.
\end{equation}
We observe that this quantum critical point is identical to that found
for the model of $M \rightarrow \infty$ infinite-range quantum rotors
\cite{YSR93, Kop94}. This was originally identified using replica
methods which are tricky for quantum phase transitions since the ratio
of the replica index to the temperature ($M/T$) must be kept fixed
while $M \rightarrow \infty$ and $T \rightarrow 0$. It was also found
using an asymptotic dynamical treatment, and it is reassuring that we
recover the same result in the limit of zero dissipation for a
treatment of a more general Hamiltonian.

\section{The phase diagram}
\label{PD}

Using $q$ and $\bar{z}$ as probes we now characterize the phase space
that is spanned by the variables $\gamma,m$ and $T$, by identifying the
coarsened and the paramagnetic regions. When the system is in the
coarsened phase, $\bar{z}=2J$ [see (\ref{z-G})], and $q \neq 0$ varies.
The $q=0$ line in the $\gamma,m,T$ phase space tells us when the
coarsened phase becomes unstable. In the paramagnetic phase $q=0$ since
the system is in equilibrium, and $\bar{z}$ varies. The spin-spin
interaction is important in the coarsened phase, and indeed $\bar{z}$
depends on $J$. In the paramagnetic phase we expect $\bar{z}$ to depend
on the two important energies $1/m$ and $T$.

\begin{figure}
  \begin{center}
    \includegraphics[width=3.25in]{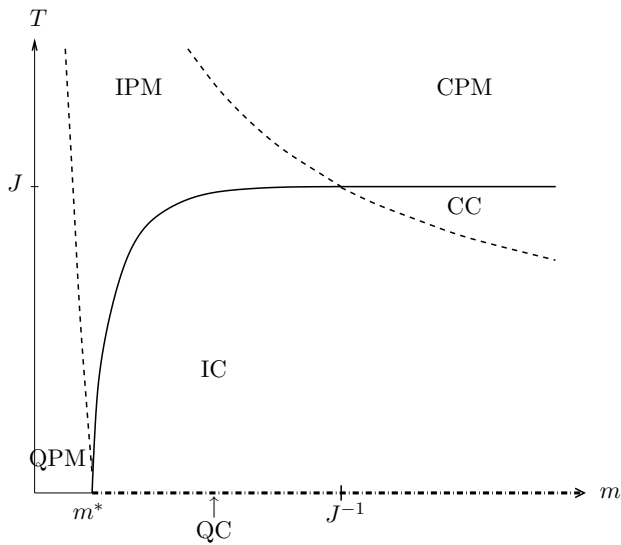}
  \end{center}
  \caption{\label{fig-B=0}The phase diagram in the $\gamma \rightarrow 0$ plane.
  The solid line depicts the second order phase transition line between the
  paramagnetic and the coarsened phases. One dashed line depicts the border
  between regions that are dominated by classical dynamics (CPM and
  CC), and the intermediate ones (IPM and IC) that are governed by both
  classical and quantum fluctuations. The other dashed line
  represents the boundary between the intermediate  paramagnetic and the quantum
  paramagnetic (QPM) regimes, where the latter is dominated by quantum dynamics.
  The quantum coarsened (QC) regime exists only
  at $T=0$, for $m>m^{*}$, the dashed-dotted line.}
\end{figure}

The system's phase behavior is presented in two sections, \ref{glass}
and \ref{PM}, associated with the coarsened and the paramagnetic phases
respectively. In the coarsened phase $\bar{z}=2J$, and we characterize
the behavior of $q$. By contrast, the system is in equilibrium in the
paramagnetic phase, so that $q=0$ and we look for $\bar{z}$. Ultimately
we show that $\bar{z}$ changes continuously from the paramagnetic value
to the coarsened one. The phase behavior in the $\gamma \rightarrow 0$
plane is displayed in Fig.~\ref{fig-B=0}, and in Fig.~\ref{fig-T=0} we
present our findings in the $T = 0$ plane. In Fig.~\ref{fig-Tvsm} we
show the transition lines in the $T-m$ plane for several values of
$\gamma$, determined by a numerical calculation. Since subsections
\ref{glass} and \ref{PM} include our calculations, they are rather
technical. For a summary of our results, we refer the reader to Tables
\ref{tab-G} and \ref{tab-PM}, which are depicted graphically in
Figs.~\ref{fig-B=0}, \ref{fig-T=0}, and \ref{fig-Tvsm}. The main
qualitative features of the system's phase behavior are summarized in
section \ref{Discussion}.

\subsection{The coarsened phase}
\label{glass}

\begin{table*}
\caption{\label{tab-G} The Edwards-Anderson parameter $q$ in the
coarsened phase, for different regions of the phase space spanned by
$m,\gamma$ and $T$; recall that here $\bar{z}=2J$. The region labels
CC, IC, and QC are according to Fig.~\ref{fig-B=0} and
Fig.~\ref{fig-T=0}}
\begin{tabular}{l|l|cr}
  \hline \hline
   & $m,\gamma$ and $T$ & $q(m,\gamma,T)$ \\ \hline
  (A) & $\begin{array}{l} \gamma \rightarrow 0 \\ T>\sqrt{J/m}
  \end{array}$ &
  $\displaystyle 1-\frac{T}{J} $ & CC\\ \hline
  (B) & $\begin{array}{l} \gamma \rightarrow 0 \\ T<\sqrt{J/m} \end{array}$ &
  $\begin{array}{c} \displaystyle 1 -
  \frac{4}{3 \pi \sqrt{mJ}}
  \left(1-x^{2} \right)^{1/2}
  \left(1+\frac{x^{2}}{2} \right)-
  \frac{2}{\pi} \frac{T}{J}
  \sin^{-1}{x}
  \\ {\rm where} \hspace{0.5cm} x= \sqrt{\frac{mT^{2}}{J}} \end{array}$
  & IC\\ \hline
  (C) & $\begin{array}{l} \gamma \rightarrow 0 \\ T=0
  \end{array}$ &
  $\displaystyle 1-\frac{4}{3 \pi \sqrt{mJ}} $ & QC\\ \hline
  (D) & $\begin{array}{l} T=0 \\ \gamma^{2} \ll mJ \end{array}$ &
  $\displaystyle 1-\frac{4}{3 \pi \sqrt{mJ}}
  \left[1 - \frac{21}{16} \frac{\gamma}{\sqrt{mJ}} -
  \frac{3}{32} \frac{\gamma^{2}}{mJ} \ln {\frac{\gamma^{2}}{mJ}}+ {\cal O}
  \left(\frac{\gamma}{\sqrt{mJ}} \right)^{2}
  \right]$ & QC\\ \hline
  (E) & $\begin{array}{l} T=0 \\ \gamma^{2} \gg mJ \end{array}$ &
  $\displaystyle 1 - \frac{1}{2 \pi \gamma}
  \ln{\frac{2 \gamma^{2}}{m J}}
  -{\cal O}\left( \frac{1}{\gamma} \right) $ & QC\\
  \hline \hline
\end{tabular}
\end{table*}

In the coarsened phase $\bar{z}=2J$, and $q$ varies; In
Table~\ref{tab-G} we summarize the expressions obtained for $q$ in the
$\gamma \rightarrow 0$ and $T=0$ planes. We identify the $q=0$ line as
a second order phase boundary, where the paramagnetic solution changes
continuously into the coarsened one, as it will be shown in section
\ref{PM}.

\subsubsection{The $\gamma \rightarrow 0$ plane}

Results (A) and (B) in Table~\ref{tab-G} are obtained for $\gamma
\rightarrow 0$, by substituting the exact expression (\ref{Gr-w-B=0})
for $\left. {\rm Im} \tilde{G}^{\rm r} \right|_{\gamma=0}$, into
Eq.~(\ref{q-approx}), and performing the integration. This part of the
phase diagram is depicted in Fig.~\ref{fig-B=0}. Expression (A)
obtained for $T>\sqrt{J/m}$ is the well-known classical result,
(\ref{q-T>}) \cite{KTJ76,CuD95a}, which depends only on the temperature
$T$. When $T=0$ and there are only quantum fluctuations, one obtains
expression (C) in Table~\ref{tab-G} for $q$ [see also (\ref{q-B=T=0})],
which depends only on $m$. Expression (B) that was obtained for
$T<\sqrt{J/m}$ has two competing terms: one that is governed by
classical dynamics and resembles the classical expression (B), and
another that is dominated by quantum fluctuations and is similar to
(C). Therefore we call this regime an intermediate region, in which
classical and quantum dynamics compete.

We identify the line $T_{\rm c}(m)$ that corresponds to $q=0$ in the
$m-T$ plane, as the phase boundary between the coarsened and the
paramagnetic phases (c.f.\ solid line in Fig.~\ref{fig-B=0}). For
$m>1/J$ the phase transition line is given by $T_{\rm c}=J$,
independent of $m$. For $m<1/J$ it is found by setting (B) equal to
$0$. The line $T_{\rm c}(m)$ is continuous approaching the point
$(m=1/J,T=J)$ from both directions. Its derivative ${\rm d}T_{\rm
c}/{\rm d}m$ is continuous at this point as well, where it becomes
zero. For $m<1/J$ the transition line terminates at the quantum
critical point $(m=m^{*},T=0)$ with an infinite derivative ${\rm
d}T_{\rm c}/{\rm d}m \rightarrow \infty$.

\subsubsection{The $T=0$ plane}

\begin{figure}
  \begin{center}
    \includegraphics[width=7cm]{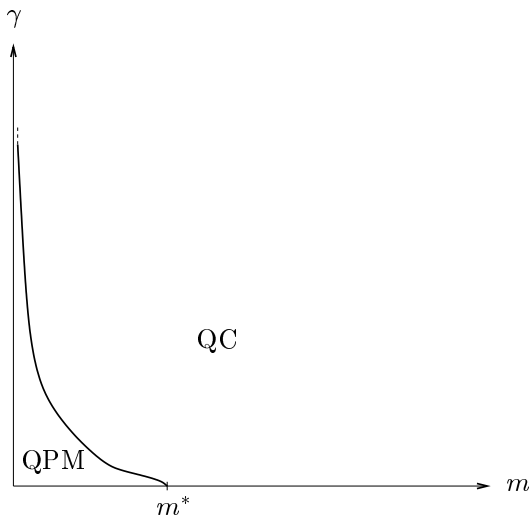}
  \end{center}
  \caption{\label{fig-T=0}The phase diagram in the $T=0$ plane. The solid line
  depicts the second order phase transition line between the
  quantum paramagnetic phase (QPM), and the quantum coarsened phase (QC).}
\end{figure}

The $T=0$ plane, depicted in Fig.~\ref{fig-T=0}, is by definition
entirely quantum. In the weak coupling limit, $\gamma^{2} \ll mJ$, we
approximate
\begin{widetext}
\begin{equation}
  \label{ImGr-B<}
  {\rm Im} \tilde{G}^{\rm r}(\omega) \simeq \left\{
  \begin{array}{lll}
  \displaystyle -{\rm Im}\left\{\frac{1+i}{J} \sqrt{\frac{\gamma \omega}{J}
  \left(1-i \frac{m \omega}{2 \gamma} \right)} \right\} +
  \frac{\gamma \omega}{J^{2}}& {\rm for} &
  \displaystyle 0 < \omega< \frac{2 \gamma}{m}, \\ \\
  \displaystyle -\frac{1}{J} \sqrt{\frac{m \omega^{2}}{J}
  \left(1-\frac{m\omega^{2}}{4 J}\right)} \left[1+ \frac{1}{2}
  \left(\frac{\gamma}{m \omega} \right)^{2} \right]+
  \frac{\gamma \omega}{J^{2}} & {\rm for} &
  \displaystyle \frac{2 \gamma}{m} < \omega < \sqrt{\frac{2J}{m}}, \\ \\
  \displaystyle -\frac{1}{J^{2}}
  \sqrt{(mJ + \gamma^{2}) \omega^{2} -
  \frac{(m \omega^{2})^{2}}{4}}+ \frac{\gamma \omega}{J^{2}}& {\rm for} &
  \displaystyle \sqrt{\frac{2J}{m}} < \omega < \sqrt{\frac{4J}{m} +
  \frac{4\gamma^{2}}{m^{2}}},
  \\ \\
  \displaystyle - \frac{2\gamma}{m^{2} \omega^{3}} & {\rm for} &
  \displaystyle \sqrt{\frac{4J}{m} + \frac{4\gamma^{2}}{m^{2}}} < \omega.
  \end{array} \right.
\end{equation}
%\end{widetext}
Substituting these expressions into Eq.~(\ref{q-Eq-T=0}) we obtain (D)
in Table~\ref{tab-G}.

The $m=0$ case is the extreme limit of strong coupling, $\gamma^{2} \gg
mJ$. In order to demonstrate that there are no coarsened solutions with
$m=0$, we substitute the imaginary part of the exact expression for
$\tilde{G}^{\rm r}(\omega)$, (\ref{Gr-w}), with $m=0$ into
Eq.~(\ref{q-Eq-T=0}). We find that
\begin{equation}
  \label{q-T=m=0}
  q=- \lim_{\omega \rightarrow \infty} \frac{1}{2 \pi \gamma}
  \ln{\frac{\gamma \omega}{J}} \rightarrow -\infty,
\end{equation}
verifying that the system must be in the paramagnetic phase when $m=0$.

In the calculation of $q$ for $\gamma^{2} \gg mJ$ and $m \neq 0$ we
approximate
%\begin{widetext}
\begin{equation}
  \label{ImGr-B>}
  {\rm Im} \tilde{G}^{\rm r}(\omega) \simeq \left\{
  \begin{array}{lll}
  \displaystyle -\frac{1}{2J^{2}} {\rm Im} \left\{
  \sqrt{2i\gamma \omega(2i \gamma \omega-4J)} \right\}& {\rm for} &
  0 < \omega<J/\gamma,
  \\ \\
  \displaystyle -\frac{1}{2 \gamma \omega} & {\rm for} &
  J/ \gamma <\omega<2B/m, \\
  \\
  \displaystyle -\frac{2 \gamma}{m^{2} \omega^{3}} & {\rm for} &
  2 \gamma /m<\omega,
  \end{array}
  \right.
\end{equation}
\end{widetext}
from which we find (E) in Table~\ref{tab-G}. The dominant logarithmic
part agrees with (\ref{q-T=m=0}). In order to find the phase boundary
for $\gamma^{2} \gg mJ$ (c.f. solid line in Fig.~\ref{fig-T=0}) we set
$q=0$ in (E) and obtain
\begin{equation}
  \label{m(B)-q=0}
  m= \frac{2 \gamma^{2}}{J} e^{-2 \pi \gamma}.
\end{equation}
Thus the phase boundary for large $\gamma$ corresponds to an
exponentially small $m$.

From Fig.~\ref{fig-T=0} it is clear that as the coupling to the bath
$\gamma$ is increased, $m$ must decrease in order to cross the
transition line. The spin-bath coupling causes a decay of the quantum
fluctuations, which in the absence of temperature, are necessary for
the system to make the transition from the coarsened to the
paramagnetic phases.

\subsubsection{Finite $T$ and $\gamma$}

\begin{figure}
  \begin{center}
    \includegraphics[width=3.25in]{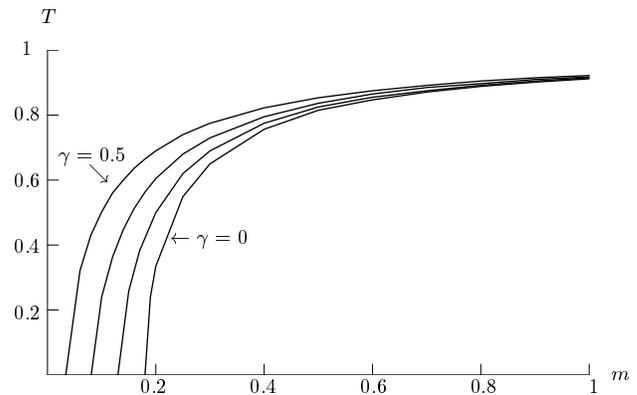}
  \end{center}
  \caption{\label{fig-Tvsm}The $q=0$ lines in the $T$ versus $m$ plane for
  $\gamma=0,0.1,0.25$ and $0.5$.}
\end{figure}

In Fig.~\ref{fig-Tvsm} we display the phase boundaries $T_{\rm c}(m)$
for a set of spin-bath couplings $\gamma$. We determined these lines
numerically, by solving $\left. q \right|_{\gamma}(m)=0$ for $T$, at a
set of $m$ values, where $q$ and $\tilde{G}^{\rm r}$ are given by
(\ref{q}) and (\ref{Gr-w}) respectively. For constant $m$ at low
temperatures, as $\gamma$ is increased, $T$ must be raised in order to
destabilize the coarsened phase. As previously discussed, the coupling
to the bath causes a decay in the quantum fluctuations. Therefore its
effect can be overcome either by decreasing $m$, thereby increasing the
quantum fluctuations, or by increasing $T$. As an aside we note that at
high temperatures, increasing $\gamma$ has a much smaller effect. Our
finding that the position of the transition line depends on the
strength of the spin-bath coupling is qualitatively similar to that
obtained in recent studies of dissipative effects on more complex
quantum systems \cite{CGL02}.

\subsection{The paramagnetic phase}
\label{PM}

\begin{table}
\caption{\label{tab-PM} The Lagrange multiplier $\bar{z}$ for different
regions of the phase space spanned by $m,\gamma$ and $T$, in the
paramagnetic phase. The region labels CPM and QPM are according to
Figs.~\ref{fig-B=0} and \ref{fig-T=0}}
\begin{tabular}{l|l|cr}
\hline \hline
  & $m,\gamma$ and $T$ & $\bar{z}(m,\gamma,T)$ &\\ \hline
  (F) & $\begin{array}{l} \gamma \rightarrow 0 \\
   \displaystyle T> \sqrt{\frac{\bar{z}+2J}{4 m}} \end{array}$ &
  $\displaystyle T\left[1+\left( \frac{J}{T}\right)^{2}\right]$ & CPM \\
  \hline
  (G) & $\begin{array}{l} \gamma \rightarrow 0 \\
   \displaystyle T<\sqrt{\frac{\bar{z}-2J}{4 m}} \end{array}$ &
  $\displaystyle \frac{1}{4m}
    \left[1 + 12 (mJ)^{2} + {\cal O}(mJ)^{4} \right] $ & QPM \\
  \hline
  (H) & $\begin{array}{l} T=0 \\ \gamma^{2} \ll mJ \end{array}$ & $\displaystyle
  \frac{1}{4m}\left[1 - \frac{8\gamma}{\pi}+12(mJ)^{2} + \cdots \right]$ & QPM
  \\ \hline
  (I) & $\begin{array}{l} T=0 \\ \gamma^{2} \gg mJ \end{array}$ &
  $\displaystyle \frac{4\gamma^{2}}{m} \exp{(-2 \pi \gamma)}$
  & QPM \\ \hline \hline
\end{tabular}
\end{table}

The expression for $\bar{z}$ that we found earlier, (\ref{z-G}),
applies only to the coarsened phase. In order to characterize the
paramagnet, we must find an expression for $\bar{z}$ in this regime.
Therefore we consider Eq.~(\ref{q-approx}) with $\tilde{G}^{\rm r}$
given by the result of Eq.~(\ref{DE-Gr-w}), substitute $q=0$, and solve
for $\bar{z}$. In Table~\ref{tab-PM} we summarize the results of this
calculation.

We remind the reader that Eq.~(\ref{DE-Gr-w}) for $\tilde{G}^{\rm r}$
applies to both the paramagnetic and the coarsened phases, where their
difference lies in their respective values of $\bar{z}$. At the
transition line, $\bar{z}=2J$ for both phases. We find that in the
paramagnet $\bar{z}$ is larger than its coarsened value of $2J$, and
that it decreases continuously to its coarsened value at the transition
line. For the cases of $\gamma \rightarrow 0$, and of $T=0$, we have
checked numerically that an attempt to solve $q=0$ using
(\ref{DE-Gr-w}) for $\tilde{G}^{\rm r}$, could be obtained beyond the
transition line (in the coarsened region) only if $\bar{z}$ continued
to decrease below the value of $2J$. Such a solution would incorporate
a finite ${\rm Im} \tilde{G}^{\rm r}(\omega)$ at $\omega=0$ which is
unphysical. Thus we deduce that the coarsened and the paramagnetic
phases coexist only at a single line, corresponding to a second order
phase transition.

Let us begin by finding $\bar{z}$ for large $T$. From
Eq.~(\ref{DE-Gr-w}) we know that $\tilde{G}^{\rm r}(0)=-
\left[\bar{z}+\sqrt{(\bar{z}^{2}-4 J^{2})}\right]/2 J^{2}$. For $T$
much larger than all other energy-scales, including $J$, and for $q=0$,
we find that $q=0=1+T \, {\rm Re}\tilde{G}^{\rm r}(0)$ [following
Eq.~(\ref{q-T>})], and therefore
\begin{equation}
  \label{z-PM-T>}
  \bar{z}=T\left(1+\frac{J^{2}}{T^{2}} \right),
\end{equation}
in agreement with the classical result \cite{KTJ76}. Note that when
$T=J$ and the $q=0$ line is approached, we obtain $\bar{z}=2J$ from
(\ref{z-PM-T>}); e.g., for the specific case of $\gamma \rightarrow 0$,
see the solid line for $m>1/J$ in Fig.~\ref{fig-B=0}. Thus $\bar{z}$,
and therefore also $\tilde{G}^{\rm r}(\omega)$ and $\tilde{G}^{\rm
s}(\omega)$ change continuously when $q=0$, and indeed $T=J$ is a
second order phase transition line.

\subsubsection{The $\gamma \rightarrow 0$ plane}
Substituting $\gamma \rightarrow 0$ in Eq.~(\ref{DE-Gr-w}) and solving
for $\tilde{G}^{\rm r}$, we find that
\begin{equation}
  \label{ImGr-B=0}
  {\rm Im}\tilde{G}^{\rm r}(\omega)=
  - \frac{1}{J} \sqrt{1 - \left(\frac{m \omega^{2} - \bar{z}}{2 J}\right)^{2}}
  \;\;\;\;\; {\rm for} \; \omega_{-} < \omega < \omega_{+},
\end{equation}
where $\omega_{\pm}=\sqrt{(\bar{z} \pm 2J)/m}$, and ${\rm Im}
\tilde{G}^{\rm r}= 0$ otherwise.

If $2T>\omega_{+}$, we use (\ref{q-approx}) to calculate $q$, and then,
substituting $q=0$ we obtain the following equation for $\bar{z}$,
\begin{equation}
  \label{Large-T}
  \frac{2T}{\pi J} \int_{\omega_{-}}^{\omega_{+}}
  \frac{{\rm d}\omega}{\omega}
  \sqrt{1 - \left(\frac{m \omega^{2} - \bar{z}}{2 J}\right)^{2}} = 1.
\end{equation}
The exact result of the integration on the left hand side is a
combination of elliptic functions, from which it is hard to isolate
$\bar{z}$. We will therefore expand in $2J/\bar{z}$. It is convenient
to make a change of variables $ \sin \theta = (m \omega^{2}-\bar{z})/2
J$. This procedure leads to expression (F) in Table~\ref{tab-PM} for
$\bar{z}$ [see also (\ref{z-PM-T>})]. Substituting this expression for
$\bar{z}$ into $\omega_{+}$ we find the line $T=\omega_{+}/2$, the
boundary between the IPM and the CPM regions above which the system
seems to be predominantly classical (c.f. dashed line in
Fig.~\ref{fig-B=0}).

For $2T<\omega_{-}$, we obtain in the same manner as above
\begin{equation}
  \label{Small-T}
  \frac{1}{\pi} \int_{\omega_{-}}^{\omega_{+}} \frac{{\rm d}\omega}{J}
  \sqrt{1 - \left(\frac{m \omega^{2} - \bar{z}}{2 J}\right)^{2}} = 1.
\end{equation}
If one neglects $2J/\bar{z}$ completely, the result is
\begin{equation}
  \label{z-B=T=J=0}
  \bar{z}=\frac{1}{4m},
\end{equation}
in agreement with \cite{CGS01}. This result is valid for $m \ll 1/8 J$.
In this case $2J/\bar{z} \ll 1$ and our approximation is consistent.

If we continue the expansion up to linear order in $2J/\bar{z}$ we
obtain expression (G) in Table~\ref{tab-PM}. Substituting $m^{*}$ into
(G) we find $\bar{z}=1.93J$ which is very close to the expected value
of $\bar{z}=2J$ in the coarsened phase. Thus, it would seem that even
close to the transition between the coarsened and the paramagnetic
states this is a good approximation. Substituting (G) into the
expression for $\omega_{-}$ we were able to draw the line
$T=\omega_{-}/2$, the dashed line between the QPM and the IPM regions
in Fig.~\ref{fig-B=0}, below which the Lagrange multiplier $\bar{z}$ is
governed by quantum behavior.

When $\omega_{-}<2T<\omega_{+}$ the system is in an intermediate regime
that is governed by both classical and quantum dynamics (IPM regime in
Fig.~\ref{fig-B=0}). We define $\theta_{T}$ such that
$\sin{\theta_{T}}=(4mT^{2}-\bar{z})/2J$. Using Eq.~(\ref{q-approx}) in
order to calculate $q$ and then setting $q=0$ we obtain
\begin{eqnarray}
  \label{Inm-T}
  \lefteqn{\frac{2T}{\pi} \int_{-\pi/2}^{\theta_{T}} {\rm d}\theta
  \frac{\cos^{2}{\theta}}{\bar{z}+2J\sin{\theta}} +} \\ \nonumber & &
  \qquad \qquad \qquad
  \frac{1}{\pi \sqrt{m}} \int_{\theta_{T}}^{\pi/2} {\rm d}\theta
  \frac{\cos^{2}{\theta}}{\sqrt{\bar{z}+2J\sin{\theta}}} =1.
\end{eqnarray}
Upon expansion in $2J/\bar{z}$, the equation for $\bar{z}$ becomes
\begin{eqnarray}
  \label{Inm-T1}
  \lefteqn{\frac{2T}{\bar{z}}
  \left\{ \! \left(\frac{\theta_{T}+\pi/2}{2 \pi}\right) \!
  \left[1 +\left(\frac{J}{\bar{z}}\right)^{2} \right] +
  \frac{2J}{3 \pi \bar{z}} \cos^{3}{\theta_{T}}
  \right\}  +} \\ \nonumber & & \!\!\!\!\!\!\!\!
  \frac{1}{\sqrt{m \bar{z}}}
  \left\{ \! \left(\frac{\pi/2-\theta_{T}}{2\pi}\right) \!
  \left[1 + \frac{3}{8}\left(\frac{J}{\bar{z}}\right)^{2} \right] -
  \frac{J}{3 \pi \bar{z}} \cos^{3}{\theta_{T}} \right\}  = 1.
\end{eqnarray}
Clearly the first term in Eq.~(\ref{Inm-T1}) describes the classical
fluctuations due to temperature, while the second describes the quantum
fluctuations due to the mass.

This point can be further clarified by rewriting Eq.~(\ref{Inm-T1}) in
the following form
\begin{equation}
  \label{Inm-T2}
  \frac{2T}{\bar{z}} \,
  f_{1} \! \left(\frac{mT^{2}}{J},\frac{\bar{z}}{J}
  \right) +
  \frac{1}{\sqrt{m \bar{z}}} \,
  f_{2} \! \left(\frac{mT^{2}}{J},\frac{\bar{z}}{J} \right)
  = 1,
\end{equation}
where $f_{1},f_{2}$ are slowly changing functions of their variables.
Neglecting terms of the order of $(J/\bar{z})^{2}$, for
$T=\omega_{+}/2$, we obtain $f_{1} \simeq 1/2$ and $f_{2}=0$, while for
$T= \omega_{-}/2$ we get $f_{1}=0$ and $f_{2} \simeq 1/2$. Solving
Eq.~(\ref{Inm-T2}) for $\bar{z}$ as if $f_{1}$ and $f_{2}$ are known we
find that
\begin{equation}
  \label{z-int}
  \bar{z} = \frac{f_{2}^{2}}{2m} + 2 f_{1} T + \frac{f_{2}}{2 \sqrt{m}}
  \sqrt{\frac{f_{2}^{2}}{m}+ 8 f_{1} T}.
\end{equation}
When $T=\omega_{+}/2$, Eq.~(\ref{z-int}) leads to $\bar{z}=T$, while
for $T=\omega_{-}/2$ one obtains $\bar{z}=1/4m$; the results for
$T>\omega_{+}/2$ and $T<\omega_{-}/2$ up to the leading order,
respectively [c.f. (F) and (G)].

\subsubsection{The $T=0$ plane}
For $T=0$ it is convenient to begin our study in the paramagnetic
region, far away from the phase boundary. Since the spin-spin
interaction is not necessary for the system to be in its paramagnetic
state, we can assume at first that $J^{2} \ll \min{\left\{4 \gamma^{2}
\bar{z}/m,\bar{z}^{2}\right\}}$, and neglect $J$ in
Eq.~(\ref{DE-Gr-w}). In this case the retarded Green function is given
by $\tilde{G}^{{\rm r}(0)}(\omega)=(m \omega^{2}+2 i \gamma \omega -
\bar{z})^{-1}$, where the superscript $0$ stands for zeroth order in
$J$. Thus we obtain from Eq.~(\ref{q-Eq-T=0}) at $q=0, T=0$ and $J=0$
the following equation for $\bar{z}$
\begin{equation}
  \label{qTJ=0}
  \frac{1}{2 \pi i \sqrt{m \bar{z} - \gamma^{2}}}
  \ln{\frac{i\gamma - \sqrt{m \bar{z} - \gamma^{2}}}
  {i\gamma+ \sqrt{m \bar{z} - \gamma^{2}}}} = 1.
\end{equation}
In the weak coupling limit, expanding Eq.~(\ref{qTJ=0}) for $\gamma^2
\ll m \bar{z}$, we find
\begin{equation}
  \label{z-TJ=0B<}
  \bar{z}=\frac{1}{4m} \left(1-\frac{8\gamma}{\pi} \right),
\end{equation}
which for $\gamma \rightarrow 0$ is consistent with
Eq.~(\ref{z-B=T=J=0}).

Now we reintroduce $J$ iteratively, by considering $\tilde{G}^{{\rm
r}(1)}=(m \omega^{2}+2i\gamma \omega-\bar{z}-J^{2}\tilde{G}^{{\rm
r}(0)})^{-1}$. We obtain from Eq.~(\ref{q-Eq-T=0}) at $q=0$ and $T=0$
the following equation for $\bar{z}$
\begin{eqnarray}
  \label{qT=0J<}
  \lefteqn{\frac{1}{4}
  \left[ \frac{1 - \frac{2}{\pi} \arctan{\frac{\gamma}{\sqrt{m(\bar{z}+J)-\gamma^{2}}}}}
              {\sqrt{m(\bar{z}+J)-\gamma^{2}}} \, + \right.}
              \\ \nonumber & & \qquad \qquad \qquad
  \left. \frac{1 - \frac{2}{\pi}
          \arctan{\frac{\gamma}{\sqrt{m(\bar{z}-J)-\gamma^{2}}}}}
              {\sqrt{m(\bar{z}-J)-\gamma^{2}}}
  \right] = 1.
\end{eqnarray}
Expanding Eq.~(\ref{qT=0J<}) in $\gamma^{2}/m \bar{z}$ and in
$J/\bar{z}$, since we expect $J$ to introduce a small correction to
$\bar{z}$ in (\ref{z-TJ=0B<}), we find (H) in Table~\ref{tab-PM}, in
agreement with (G) for $\gamma \rightarrow 0$.

Returning to Eq.~(\ref{qTJ=0}) and assuming strong coupling so that
$\gamma^{2} \gg m \bar{z}$, we obtain (I), which will equal $2J$ when
$m=2\gamma^{2}/J \exp{(-2 \pi \gamma)}$. This is exactly the same
expression for the transition line that we obtained for the coarsened
phase [see (\ref{m(B)-q=0})]. Thus for large $\gamma$ the transition is
also continuous.

\section{Discussion} \label{Discussion}

In this paper we present a study of the phase behavior of the
disordered quantum $p=2$ spherical model as a function of temperature,
dissipation (spin-bath coupling) and strength of quantum fluctuations
(mass) at long times, identifying different paramagnetic and coarsened
phase regions. Using the weak-ergodicity breaking ansatz and a linear
spin-bath coupling, we characterize the system's behavior in the
stationary regime. We do so by exploiting the fact that its retarded
Green function is independent of its statistical counterpart, and is
time-translationally invariant at long-time scales. In doing so, we
demonstrate analytically that in the stationary regime the statistical
and the retarded Green functions are linked by the
fluctuation-dissipation theorem (FDT).

We use the spherical constraint and the fluctuation-dissipation theorem
in the stationary regime to characterize the phase behavior of the
system. More specifically, the long-time behavior of $\bar{z}$, the
Lagrange multiplier that enforces the spherical constraint, is crucial
for distinguishing different paramagnetic regimes. Similarly the
Edwards-Anderson order parameter, $q$, determined from the stationary
statistical Green function (FDT) and matching conditions at equal-times
(spherical constraint), characterizes different coarsened phases. We
find that all the boundaries between the paramagnetic and the coarsened
regions, determined by the $q=0$ lines, in the $\gamma \rightarrow 0$
and the $T=0$ planes are continuous; this is in marked contrast to the
situation for quantum $p \geq 3$ spherical models where quantum
fluctuations at $T=0$ drive the transition first-order
\cite{CGS00,CGS01,BiC01}. Furthermore, we identify a quantum critical
point for $T=0$ and $\gamma \rightarrow 0$ that coincides with the one
found for a model of infinite-range, infinite-component quantum rotors
in the absence of dissipation \cite{YSR93, Kop94}.

The Edwards-Anderson order parameter is used to probe the different
types of coarsened phases; here $\bar{z}$ is a constant ($\bar{z}=2J$).
More specifically, the behavior of $q$ determines whether the system's
dynamics are dominated by classical or by quantum energy-scales. From
our results, it is clear that the increase of either quantum or thermal
fluctuations (c.f. Figs.~\ref{fig-B=0} and \ref{fig-T=0}), by raising
the inverse-mass or the temperature respectively, leads to the
destabilization of the coarsened phase. Furthermore an increase of the
spin-bath coupling, $\gamma$, causes a decay of the quantum
fluctuations; the inverse-mass must be enhanced still further to
approach the phase boundary. Finally for constant mass and increasing
$\gamma$ at low temperatures, an increase in $T$ is necessary to cross
the $q=0$ phase boundary (c.f. Fig.~\ref{fig-Tvsm}); the effect of the
spin-bath coupling is significantly reduced at higher $T$.

In the paramagnetic phase, $q = 0$, and the system is characterized by
its long-time Lagrange multiplier, $\bar{z}$, which enforces the
spherical constraint. In both the $\gamma \rightarrow 0$ and the $T=0$
planes (c.f. Figs.~\ref{fig-B=0} and \ref{fig-T=0}, respectively) a
quantum paramagnetic phase exists in the limit of vanishing mass.
However in the $\gamma \rightarrow 0$ case there are crossovers between
different paramagnetic regions as both temperature and mass are
increased, corresponding to states where classical energy-scales assume
increasing importance compared to their quantum counterparts.

The simplicity of this quantum $p=2$ spherical model is both its
blessing and its curse. It is exactly this feature that permits us to
show explicitly the validity of the fluctuation-dissipation theorem in
the stationary regime; furthermore this system is accessible to
analytic study in the limit of vanishing dissipation. However, we
should also point out that here the response is {\em always}
independent of temperature, a feature that makes it particularly
awkward for comparison with experiment in the classical and quantum
regimes. Furthermore we do not have any spatial information, so
possible overlap with real coarsening measurements is rather difficult.
However some dialogue between theory and experiment may be possible
near the quantum critical point, where issues of length-scales assume
reduced importance. As a first step towards making such contact with
experiment, we need to establish whether the quantum critical point
studied here is in the same universality class as that of the Ising
model in a transverse magnetic field. It is intuitive to relate the
latter to the $M=1$ quantum rotor case \cite{YSR93,RSY98,KoP97}, which
would suggest that it is in a different universality class from the
$p=2$ quantum spherical model (that strongly resembles the $M
\rightarrow \infty$ case) studied here. However the distinction between
continuous and discrete spin symmetries may be important \cite{KoP97},
so this argument is inconclusive. On another note, it is known that one
can glean spatial information from the study of the classical $p=2$
spherical model by exploiting a mapping on long time-scales to the
three dimensional $O(n)$ model \cite{CuD95a,BBM96}, and an analogous
mapping should be explored for the quantum system. Finally we plan to
characterize the dynamics of this simple system near its quantum
critical point in order to study whether it has any slow relaxational
modes that are not shared by its classical counterpart.

{\em Note added in proof}: Recently, we learned of earlier studies
\cite{SRO95-SeG95} of the critical behavior of quantum rotors at finite
dissipation. In the relevant parameter regime our results are
qualitatively similar. We thank S. Sachdev and A. Sengupta for bringing
these papers to our attention.

\begin{acknowledgements}

We would like to thank G. Aeppli, G. Biroli, J. Brooke, L.~B. Ioffe, I.
Lerner, O. Parcollet, and T.~F. Rosenbaum for a number of insightful
discussions. This work was partially done at NEC Research Institute. We
also acknowledge support from NSF grant 4-21262.
\end{acknowledgements}

\end {document}